# Realization of air-stable two-dimensional superconductor $Nb_2Pd_3Te_5$ with quasi-one-dimensional pair density modulation

Jiayi Wang[1,2#], Hui Guo[1,2#*], Hao Zhang[1,2#], Haowei Chen[3#], Peixuan Li[1,2], Xianghe Han[1,2], Ziang Wang[1,2], Siyu Xu[1,2], Qian Fang[1,2], Haohao Xu[1,2], Shixuan Du[1,2], Chengmin Shen[1,2], Hui Chen[1,2*], Wang Yao[3], and Hong-Jun Gao[1,2*]

[1] Beijing National Center for Condensed Matter Physics and Institute of Physics, Chinese Academy of Sciences, Beijing 100190, PR China

[2] School of Physical Sciences, University of Chinese Academy of Sciences, Beijing 100190, PR China

[3] New Cornerstone Science Laboratory, Department of Physics, The University of Hong Kong, Hong Kong, PR China

[#]These authors contributed equally to this work

[*]Correspondence to: hjgao@iphy.ac.cn, hchenn04@iphy.ac.cn, guohui@iphy.ac.cn

**Two-dimensional (2D) superconductors provide a fertile platform for exploring reduced-dimensional superconductivity and emergent quantum phenomena. Incorporating quasi-one-dimensional (quasi-1D) structural motifs into 2D superconductors offers a powerful route to engineer strong electronic anisotropy, enabling unconventional superconducting states and anisotropic superconducting transport functionalities. However, such systems remain rarely realized. Here we report the realization of a 2D superconductor $Nb_2Pd_3Te_5$, exhibiting an intrinsic quasi-1D pair density modulation. Monolayer and bilayer $Nb_2Pd_3Te_5$ is synthesized via van-der-Waals epitaxy. Using ultralow-temperature scanning tunneling microscopy/spectroscopy, we observe the quasi-1D crystal structure and superconductivity below ~0.6 K with a pronounced quasi-1D pair density modulation. Remarkably, both monolayer and bilayer $Nb_2Pd_3Te_5$ show strong air stability. Our findings establish atomically 2D $Nb_2Pd_3Te_5$ as a robust and promising platform for exploring novel low-dimensional quantum phenomena and anisotropy-enabled superconducting devices.**

## Introduction

Two-dimensional (2D) superconductivity has attracted great interest as a fertile platform for exploring emergent low-dimensional quantum phenomena[1–8]. In particular, 2D van-der-Waals (vdW) superconductor offers unique advantages by combining reduced dimensionality with atomically-thin, weakly coupled layered structures, enabling unconventional superconducting states and device functionalities inaccessible in bulk systems[9–12]. Notable examples include Ising superconductivity intertwined with charge density wave (CDW) order in monolayer $NbSe_2$ and $TaS_2$[13–16], as well as pair density modulation states in monolayer $MoTe_2$[17] and iron-based superconductors[18,19] , highlighting remarkable tunability of superconducting states in 2D limit. Moreover, their atomically clean surfaces and weak interlayer coupling facilitate controlled stacking and seamless integration into designed architectures[20–23], making 2D vdW superconductors ideal building blocks for Josephson junctions, proximity and gate-tunable superconducting devices[24–26].

Despite these advantages, there are two major demands for realizing functional 2D vdW superconductors. First, incorporating quasi-one-dimensional (quasi-1D) structural motifs into 2D superconductors offers an effective route to introduce strong anisotropy, enabling even richer emergent quantum phenomena[27–29] and new opportunities for anisotropic superconducting functionalities[30,31]. Second, many atomically-thin superconductors are highly sensitive to ambient conditions, posing significant obstacles for device fabrication and practical applications[6,32]. Developing air-stable 2D superconducting materials is therefore essential for advancing both fundamental studies and device engineering[33,34]. To date, extensive efforts have been devoted to quasi-1D superconductors, unveiling phenomena such as anisotropic vortex textures[35,36], intertwined CDW and superconductivity[37], quantum Griffiths singularities[38], and possible topological superconducting states[39]. All previous studies of quasi-1D structure and superconductivity have thus far been largely limited to three-dimensional bulk crystals. In particular, bulk/polycrystalline $Nb_2Pd_3Te_5$ has recently been reported as a new quasi-1D superconductor exhibiting a superconducting transition near 3.3 K and pronounced anisotropic properties[40]. However, whether superconductivity survives in the atomically thin limit, and how dimensional reduction reshapes its quasi-1D electronic anisotropy and superconducting state, remain unknown.

Here, we report the experimental realization of an air-stable, atomically-thin quasi-1D superconductor

$Nb_2Pd_3Te_5$. By employing vdW epitaxy, we synthesize $Nb_2Pd_3Te_5$ down to single atomic layers. Using ultralow-temperature scanning tunneling microscopy/spectroscopy (STM/STS), we observe stripe-like quasi-1D crystal structure and pronounced anisotropic electronic modulation. In addition, we reveal the emergence of superconductivity below ~0.6 K, accompanied by a pronounced quasi-1D modulation of superconducting pair density. Remarkably, monolayer $Nb_2Pd_3Te_5$ shows no detectable structural degradation after prolonged exposure to ambient conditions, establishing it as a rare example of an air-stable low-dimensional superconductor. Our work identifies 2D $Nb_2Pd_3Te_5$ as a robust platform for exploring emergent low-dimensional and anisotropic quantum states of superconductors, as well as functional superconducting devices.

## Results and discussion

$Nb_2Pd_3Te_5$ crystallizes in an orthorhombic structure[40] (space group *Pnma*, No. 62) with lattice constants $a$ = 1.47 nm, $b$ = 0.36 nm, and $c$ = 1.87 nm. Its crystal structure consists of quasi-1D Nb-Pd-Te chains aligned along $c$-axis, which alternate in up- and down-orientated configurations along $c$-axis and are stacked via weak vdW coupling along $a$-axis, forming a quasi-1D lattice framework (Figure 1a). Based on arrangements of the topmost Te atoms, the two alternating oriented quasi-1D chains are denoted as T-chain and H-chain (Figure 1a).

Atomically-thin $Nb_2Pd_3Te_5$ is synthesized on HOPG (highly oriented pyrolytic graphite) substrate by vdW epitaxy. Sharp low-energy electron diffraction (LEED) patterns indicate the high crystalline quality and well-defined in-plane periodicity of the as-grown $Nb_2Pd_3Te_5$ (Figure 1b). Large-scale STM topography shows atomically flat terraces corresponding to monolayer (1L) and bilayer (2L) $Nb_2Pd_3Te_5$ (Figure 1c). By tuning the growth conditions, the thickness of $Nb_2Pd_3Te_5$ films can be well controlled (Figure S1). The line profiles quantitatively reveal a height of ~1.0 nm for the monolayer $Nb_2Pd_3Te_5$ on graphene substrate, exceeding the interlayer spacing of ~0.70 nm, suggesting weak substrate coupling and quasi-freestanding electronic characteristics of the epitaxial films (Figure S1). Zoom-in STM image clearly shows a uniform stripe-modulated surface with a periodicity of ~1.87 nm perpendicular to the stripe (Figure 1d, S2), consistent with the lattice constant along $c$-axis and reflecting quasi-1D structural nature of the $Nb_2Pd_3Te_5$. Such stripe-like topographic modulation is visible across different STM images and consistently observed throughout the sample, and the stripe orientation is reproducibly locked to the crystallographic axes across different regions, reflecting its intrinsic quasi-1D characteristics. Atomically-resolved STM image further resolves the surface atomic structure with a lattice periodicity of ~0.36 nm along *b*-axis (Figure 1e, S2, S3). Although the long-wavelength stripe modulation becomes less visually prominent in atomic-resolution STM images because the contrast is dominated by atomic corrugation, the stripe structure remains consistently present and can be more clearly resolved under different imaging conditions (Figure S3). Moreover, the simulated STM image based on the structural model in Figure 1a shows excellent agreement with the experimental observations, unambiguously confirming the formation of quasi-1D $Nb_2Pd_3Te_5$ films.

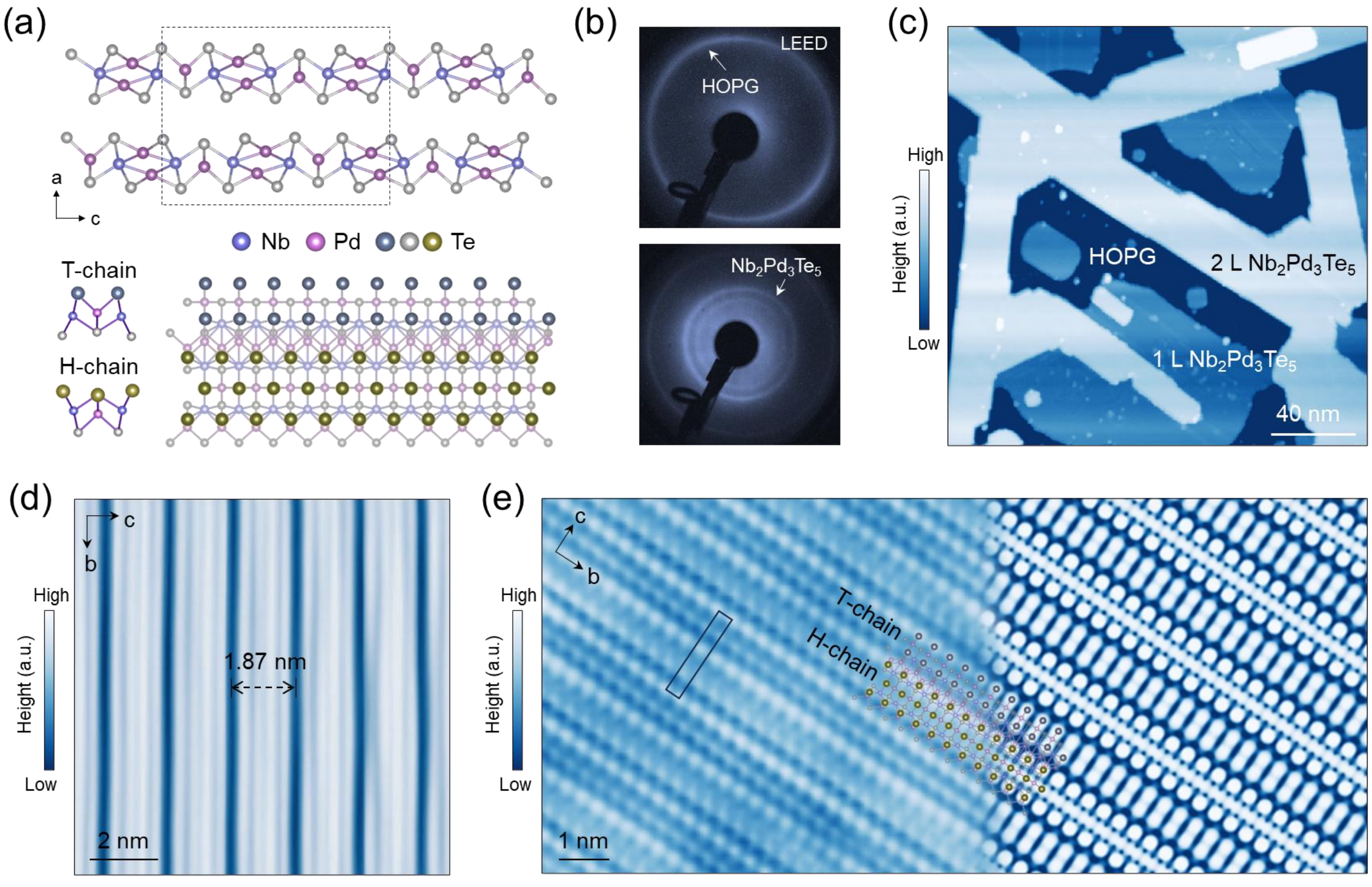

**Figure 1. Synthesis of monolayer and bilayer $Nb_2Pd_3Te_5$ on HOPG substrate.** (a) Side and top views of the crystal structure of $Nb_2Pd_3Te_5$, showing a vdW layered structure composed of quasi-1D atomic chains. Within a single unit cell, the quasi-1D chains exhibit two distinct surface atomic configurations, classified as tetragonal (T-chain) and hexagonal (H-chain) according to the arrangement of the topmost Te atoms. The dashed black rectangle denotes the unit cell. (b) LEED patterns for the cleaved HOPG substrate and as-grown $Nb_2Pd_3Te_5$ sample, showing sharp diffraction patterns. (c) Large-scale STM topography of the epitaxial 2D $Nb_2Pd_3Te_5$ grown on HOPG substrate ($V_s = -2$ V, $I_t = 50$

pA), showing coexisting both monolayer and bilayer regions. (d) High-resolution STM image of the $Nb_2Pd_3Te_5$ ($V_s$ = −100 mV, $I_t$ = 200 pA), showing pronounced quasi-one-dimensional stripe features, with a periodicity of ~1.87 nm across the stripe. (e) Atomically-resolved STM image of the $Nb_2Pd_3Te_5$ ($V_s$ = −50 mV, $I_t$ = 2 nA), showing both H-chain and T-chain motifs within a single unit cell (black rectangle), which is in good agreement with the simulated STM image based on the atomic structure in (a).

We next study the electronic properties of the 2D vdW $Nb_2Pd_3Te_5$ by using low-temperature STM/STS. Figure 2a displays an STM topography containing adjacent 1L and 2L $Nb_2Pd_3Te_5$ regions separated by a straight, well-defined step edge. Spatially-averaged d$I$/d$V$ spectra taken at H-chain and T-chain for both thickness exhibit four pronounced characteristic peaks, labeled as $P_1$-$P_4$ (Figure 2b). The electronic spectra obtained from the same chain type in the monolayer and bilayer are nearly identical, suggesting that interlayer coupling has a negligible influence on the electronic structure of the $Nb_2Pd_3Te_5$. In contrast, pronounced spectral differences emerge between the two inequivalent chains. In particular, the spectral weight associated with peaks $P_2$ (~300 meV) and $P_4$ (~770 meV) is strongly enhanced on the T-chains but markedly suppressed on the H-chains. The d$I$/d$V$ linecuts along both T- and H- chains reveal spatially uniform electronic states (Figure S4), whereas it reveals pronounced modulations perpendicular to the chain direction (Figure 2c). Consistently, spatially-resolved d$I$/d$V$ maps acquired at −200 mV and +200 mV (Figures 2d-f and S5) show that while the local density of states (LDOS) at occupied energies is nearly uniform, while the unoccupied states are strongly localized on the H-chains. To quantify this energy-dependent LDOS modulation, we performed a cross-correlation analysis using the +200 mV d$I$/d$V$ map as the reference. The correlation coefficient is weak at negative bias voltages and near zero bias, but increases rapidly in the positive-bias regime and approaches unity toward +200 mV (Figure S6), confirming that the unoccupied states exhibit a highly correlated quasi-1D spatial distribution. These spatially periodic modulations of the electronic states, following the alternating arrangement of T- and H-chains, directly reflect the intrinsic electronic anisotropy and quasi-1D character of $Nb_2Pd_3Te_5$. The chain-dependent LDOS modulation observed here may resemble electronic textures commonly discussed in quasi-1D systems with CDW correlations. However, the observed modulation follows the intrinsic lattice periodic arrangement of the inequivalent H- and T-chains, without additional superlattice periodicity expected for a long-range CDW order. Consistently, the LEED measurements do not show extra diffraction spots associated with long-range CDW modulation. Furthermore, d$I$/d$V$ linecut taken across the step edge between the monolayer and bilayer regions exhibit a distinct localized state near the interface (Figure 2g,h). Such interfacial states likely arise from local structural or electronic reconstruction induced by thickness discontinuity, reflecting the sensitivity of the quasi-1D electronic system to dimensionality and coordination perturbations in the epitaxial $Nb_2Pd_3Te_5$[41,42]

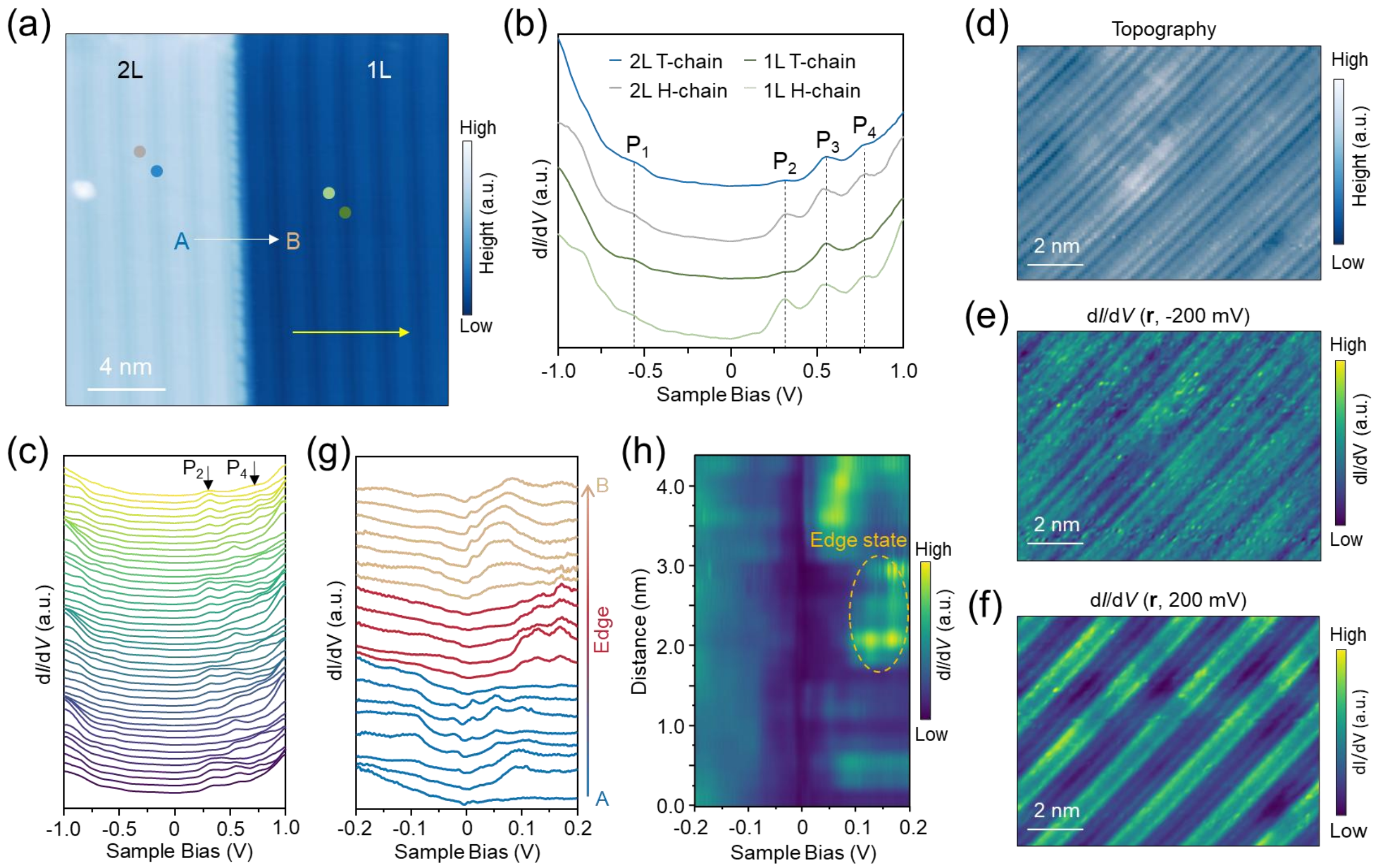

**Figure 2. Observation of quasi-1D structure and electronic properties of $Nb_2Pd_3Te_5$.** (a) STM topography, showing coexisting 1L and 2L $Nb_2Pd_3Te_5$ regions separated by a sharp step edge. (b) Spatially-averaged d*I*/d*V* spectra acquired on the H-chains and T-chains for both 1L and 2L $Nb_2Pd_3Te_5$ (positions marked by solid circles in (a)), respectively, showing nearly identical overall electronic properties for the monolayer and bilayer, characterized by four pronounced peaks labeled $P_1$-$P_4$. The electronic states associated with $P_2$ and $P_4$ display notable chain-dependent modulation between the H- and T-chains. (c) The d*I*/d*V* linecut taken along the yellow arrow in (a), showing periodic modulation of the electronic states at $P_2$ and $P_4$, indicative of quasi-1D electronic behavior. (d-f) STM topography (d) and corresponding d*I*/d*V* maps acquired at −200 mV (e) and +200 mV (f), showing that the electronic states near +200 meV are strongly localized at H-chains, further confirming the quasi-one-dimension nature of the electronic structure. (g,h) The d*I*/d*V* linecut and corresponding intensity map taken across the step edge (along the white arrow in (a)), showing a distinct edge state at the 1L-2L interface. All STM/STS data were acquired at approximately 6 K.

We further investigate the superconducting properties of the 2D $Nb_2Pd_3Te_5$ using ultra-low-temperature STM/STS measurements. A representative differential conductance spectrum acquired on bilayer $Nb_2Pd_3Te_5$ at 5 mK reveals a pronounced V-shaped superconducting gap with symmetric coherence peaks centered at the Fermi level (Figure 3a). The V-shaped curve is well reproduced by a Dynes fit, yielding a superconducting gap Δ~121 μeV. The well-defined gap symmetry and sharp quasiparticle coherence peaks indicate a clean superconducting condensate and high sample homogeneity[43,44]. A series of temperature-dependent d*I*/d*V* spectra shows that the coherence peaks gradually weaken and the superconducting gap closes upon increasing temperature from 10 mK to 614 mK, confirming the superconducting origin of the observed gap (Figure 3b). Each spectrum at different temperatures can be consistently fitted by the Dynes function using a single fitting parameter Δ(T), indicating an intrinsic superconducting DOS (Figure 3c). The extracted gap Δ(T) follows the Bardeen-Cooper-Schrieffer (BCS) behavior, yielding a superconducting critical temperature $T_c$ of approximately 613 mK. The corresponding ratio $2\Delta_0/k_BT_c \sim 4.6$ slightly exceeds the weak-coupling BCS value of 3.53, suggesting moderately strong electron–phonon coupling in this quasi-1D system[45,46]. The superconducting transition temperature observed in the present atomically thin $Nb_2Pd_3Te_5$ is substantially lower than that reported previously in bulk/polycrystalline samples[40]. Such suppression may originate from reduced dimensionality, which generally can weaken the interlayer coupling, electronic structure and superconducting phase coherence.

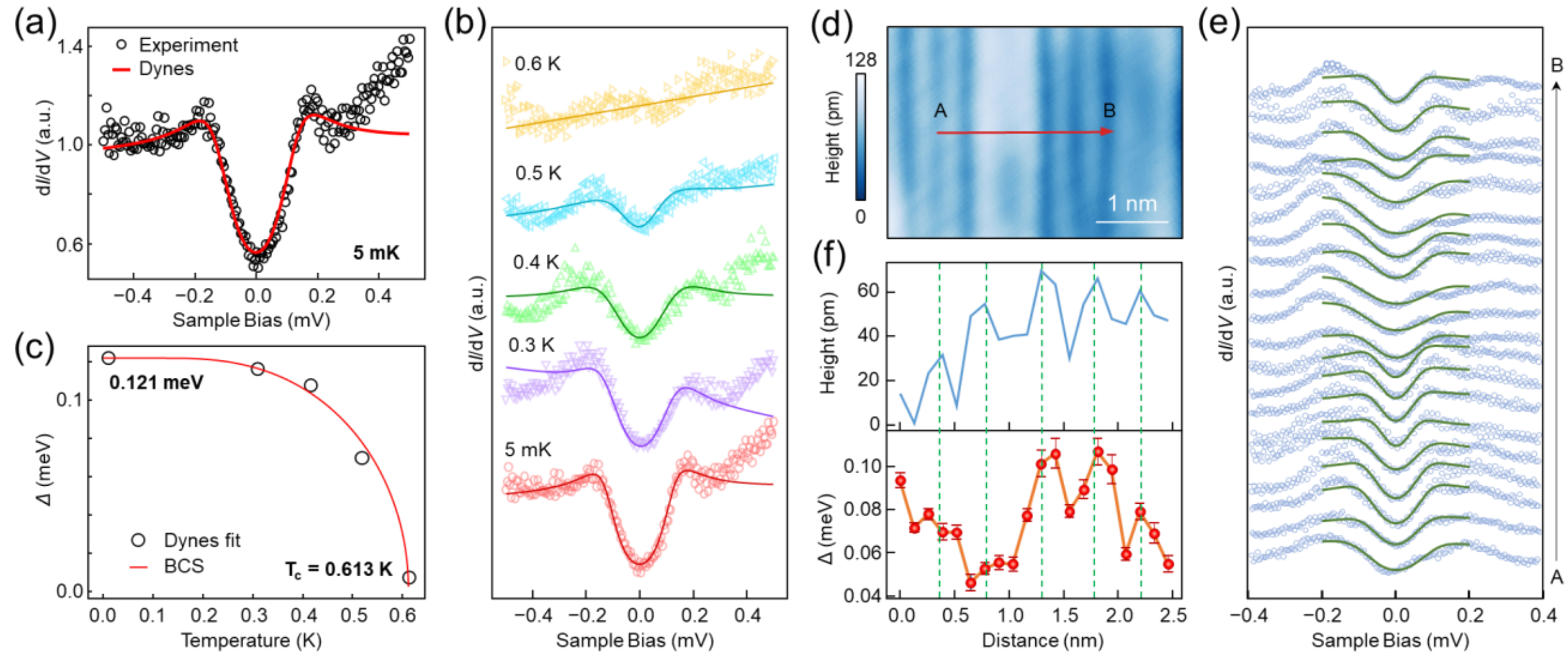

**Figure 3. Superconductivity and quasi-1D pair density modulation of the bilayer $Nb_2Pd_3Te_5$.** (a) Low-temperature d*I*/d*V* spectrum of bilayer $Nb_2Pd_3Te_5$ at 5 mK, showing a clear V-shaped superconducting gap with symmetric coherence peaks. The red curve is a Dynes-function fit, yielding an energy gap $\Delta \sim 121$ μeV. (b) Temperature-dependent d*I*/d*V* spectra from 10 mK to 614 mK, where the gap gradually narrows and coherence peaks weaken with increasing temperature, consistent with BCS-type behavior. (c) Extracted gap Δ as a function of temperature with the BCS fit (red line) gives $T_c \sim 613$ mK. (d) STM image showing a unidirectional stripe. (e) The d*I*/d*V* linecut along the red arrow in (d) with Dynes-function fit (red curves), showing the spatial variations in superconducting gap. The d*I*/d*V* spectra is normalized by the d*I*/d*V* value at -0.4 meV to exclude the possible tip–sample convolution or local geometric effects. (f) The line profile of height (upper) and superconducting gap (lower) subtracting from the linecut in (e), showing the pair density modulation on the stripe of bilayer $Nb_2Pd_3Te_5$. The green dashed lines represent the modulation of quasi-1D chains. The error bars were generated from the BCS fitting of superconducting gap.

Strikingly, the superconductivity in the 2D $Nb_2Pd_3Te_5$ exhibits pronounced spatial quasi-1D modulations. Figure 3e presents a spatially resolved d*I*/d*V* linecut acquired across H- and T-chains along the direction indicated in the STM topography shown in Figure 3d. Well-defined superconducting gaps with symmetric coherence peaks are observed at all spatial locations along the linecut, demonstrating that superconductivity in bilayer $Nb_2Pd_3Te_5$ is globally robust. Nevertheless, the spatial variations in both the superconducting gap magnitude and coherence peak intensity are clearly resolved. To quantitatively characterize this spatial variation, each spectrum in the linecut is fitted by the Dynes model, allowing extraction of the local superconducting gap as a function of position. The resulting superconducting gap Δ exhibits a pronounced modulation along the *c*-axis, closely following the underlying quasi-1D structural modulation (Figure 3f), with the gap maxima spatially correlated with the topographic maxima and the gap minima aligned with the topographic minima. It should be noted we did not observe the similar pair density modulation in another superconductor $NbTe_2$ which has comparable superconducting gap size and quasi-1D topographic feature (Figure S7)[36]. Thus, this observation provides direct evidence for the formation of a quasi-1D pair density modulation state in the bilayer $Nb_2Pd_3Te_5$.

Although no distinct CDW-like gap feature is resolved within the present STS measurements, the pronounced quasi-1D electronic anisotropy may still provide an important electronic environment for the emergence of the superconducting gap modulation observed here. More broadly, such spatially modulated superconductivity connects $Nb_2Pd_3Te_5$ to a wider class of anisotropic superconducting systems in two dimensions. For example, superconductivity in two-dimensional electron gases at $EuO/KTaO_3(111)$ interfaces is accompanied by pronounced in-plane transport anisotropy and a stripe-like phase when global superconductivity is weakened by temperature or magnetic field, suggesting an incipient or fluctuating pair density wave state[47]. In contrast to such interfacial electron-gas systems, $Nb_2Pd_3Te_5$ represents an intrinsic quasi-1D vdW superconductor in which the electronic anisotropy originates directly from the inequivalent H- and T-chain lattice structure. Importantly, the superconducting gap modulation is directly visualized by ultralow-temperature STM/STS in real space and is spatially correlated with the underlying stripe-like lattice modulation. These results establish atomically thin $Nb_2Pd_3Te_5$ as a promising platform for exploring anisotropic and spatially modulated superconductivity in the two-dimensional limit. Additionally, future transport measurements, including possible anisotropic transport, critical current behavior, and Berezinskii–Kosterlitz–Thouless physics, may provide further insight into how the local superconducting modulation observed here evolves into macroscopic superconducting coherence in this quasi-1D system.

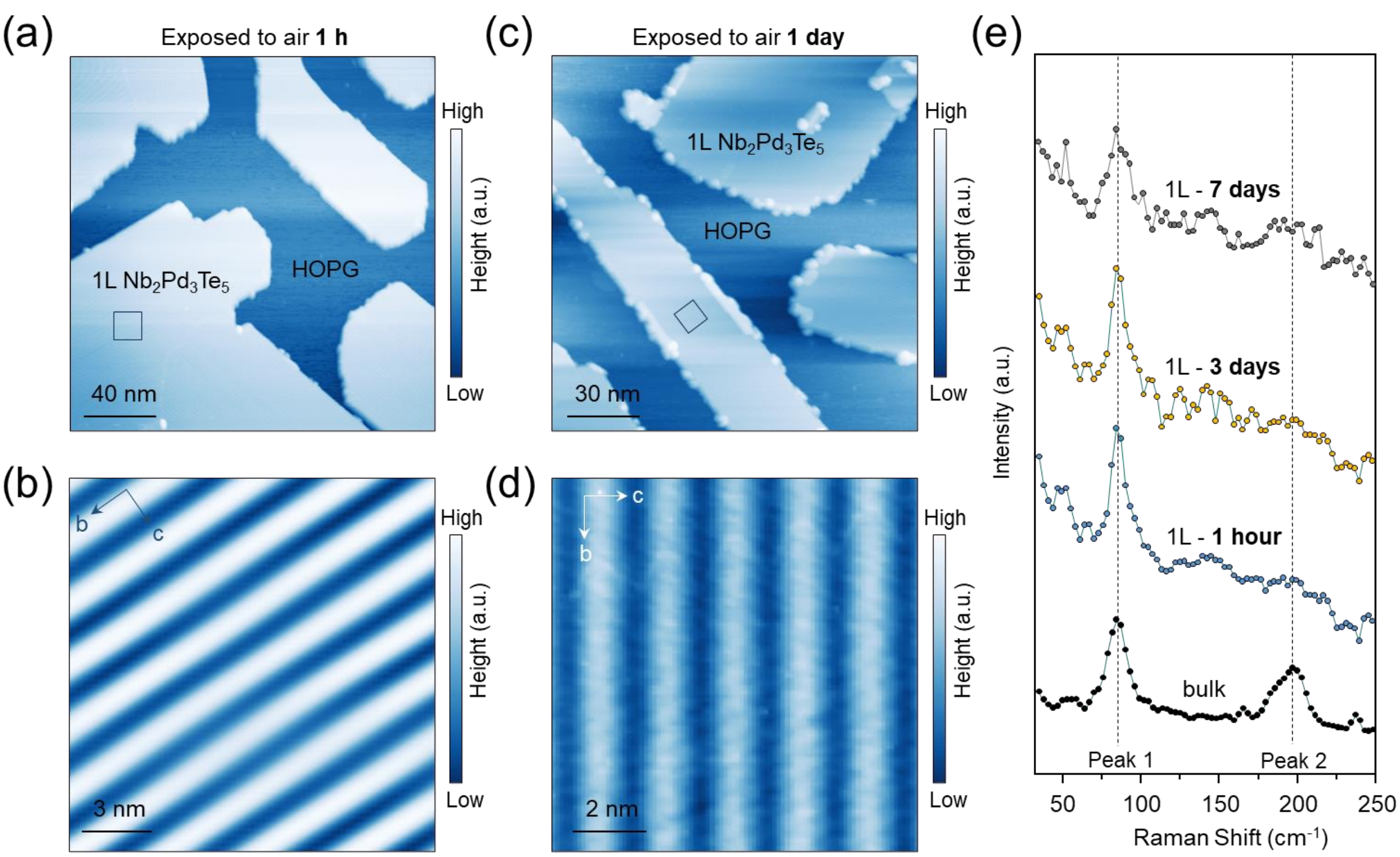

**Figure 4. Stability of the monolayer $Nb_2Pd_3Te_5$ in ambient conditions.** (a) Large-scale STM topography of monolayer $Nb_2Pd_3Te_5$ after exposure to ambient air for 1 hour, an atomically clean surface without observable

adsorbates or structural defects. (b) Zoom-in STM image taken from the region marked by the black box in (a), showing stripe features identical to those of the as-grown film. (c) Large-scale STM topography of monolayer $Nb_2Pd_3Te_5$ after exposure to ambient air for 1 day, showing a largely clean surface with only minor adsorbates appearing near the edges. (d) Zoom-in STM image taken from the region marked by the black box in (c), showing the intact stripe structure, indicating negligible structural degradation. (e) Raman spectra of the bulk $Nb_2Pd_3Te_5$ and as-grown monolayer after exposure to ambient air for 1 hour, 1 day, 3 days, and 7 days, respectively. Bulk $Nb_2Pd_3Te_5$ exhibits two characteristic vibrational modes at approximately 83 $cm^{-1}$ (Peak 1) and 193 $cm^{-1}$ (Peak 2), whereas the monolayer mainly displays peak 1 due to the suppression of out-of-plane vibration in the two-dimensional limit. The persistence of Peak 1 upon prolonged air exposure confirms the robust ambient stability of the monolayer $Nb_2Pd_3Te_5$.

Finally, we expose monolayer $Nb_2Pd_3Te_5$ to air at room temperature to assess the ambient stability of the 2D $Nb_2Pd_3Te_5$, as chemical stability is a critical property for materials used in electronic applications[48], especially for atomic-thickness 2D superconductors. After 1 hour of air exposure, large-scale STM topography reveals that the monolayer $Nb_2Pd_3Te_5$ islands remain clean and smooth, without obvious impurities or surface disorders (Figure 4a). The zoom-in STM image shows stripe modulations identical to those of the as-grown samples, indicating the absence of noticeable structural degradation (Figure 4b). Notably, even after one day of ambient exposure, only a small number of adsorbates are observed, predominantly localized near the edges of the monolayer $Nb_2Pd_3Te_5$ islands (Figures 4c). In contrast, the interior regions remain atomically ordered, with periodic lattice structure fully preserved (Figures 4d). To further corroborate the excellent ambient stability, we perform Raman spectroscopy on monolayer $Nb_2Pd_3Te_5$ samples after prolonged air exposure (Figure 4e). Bulk $Nb_2Pd_3Te_5$ crystals exhibit two characteristic Raman modes at 83 $cm^{-1}$ and 193 $cm^{-1}$. Compared with the bulk sample, the Raman intensity of the 193 $cm^{-1}$ mode is strongly suppressed in the monolayer limit, possibly because this phonon mode contains a relatively stronger out-of-plane vibrational component[49-51]. Importantly, the Raman measurements here primarily serve as a probe of environmental stability. The nearly unchanged Raman peak position and intensity of the monolayer $Nb_2Pd_3Te_5$ after prolonged air exposure indicate its robust structural and chemical stability under ambient conditions. We note that the present measurements mainly establish the environmental robustness of the material framework. Direct spectroscopic verification of superconductivity after air exposure remains challenging because of the extremely small superconducting energy scale in this system and will require further dedicated investigations. Nevertheless, these results strongly demonstrate that 2D $Nb_2Pd_3Te_5$ exhibits exceptional ambient stability, surpassing that of most reported low-dimensional superconductors. This chemically robust vdW superconductor therefore provides a highly promising platform for future electronic and quantum device applications.

## Conclusions

We have synthesized a 2D vdW layered superconductor $Nb_2Pd_3Te_5$, featuring an intrinsic quasi-1D lattice structure and pronounced quasi-1D electronic modulations. Ultra-low-temperature STM measurements evidence the superconductivity in the atomically-thin $Nb_2Pd_3Te_5$, and strikingly, reveal a quasi-1D pair density modulation state that is closely correlated with the underlying stripe-like structural motif. In addition, 2D $Nb_2Pd_3Te_5$ exhibits exceptional ambient stability, distinguishing it from most reported 2D vdW superconductors. Our findings establish $Nb_2Pd_3Te_5$ as a robust and versatile platform for exploring low-dimensional quantum phenomena and electronic device applications.

## Methods

**Sample preparation.** Monolayer and bilayer $Nb_2Pd_3Te_5$ were epitaxially grown on HOPG substrates by molecular beam epitaxy (MBE) under ultra-high-vacuum (UHV) conditions (base pressure ~$5\times10^{-10}$ mbar). High-quality HOPG substrate with atomically flat surfaces was obtained by mechanically exfoliation followed by degassing at ~870 K for several hours. The 2D $Nb_2Pd_3Te_5$ were grown by co-evaporating Nb (99.9%, Goodfellow Cambridge Ltd.), Pd (99.95%, Goodfellow Cambridge Ltd.), and Te (99.99%, Sigma-Aldrich) atoms with the substrate maintained at 673 K. During growth, the Te flux is kept approximately one order of magnitude higher than the Nb and Pd fluxes to ensure a Te-rich atmosphere.

**STM/STS, LEED characterizations.** After the sample growth, it was then transferred to an STM system via a home-made UHV suitcase. STM/STS measurements were performed in a UHV ultra-low-temperature STM system (base pressure ~$1\times10^{-10}$ mbar). During STM/STS measurements, the sample temperature is stabilized at 6 K and could be further reduced to 5 mK using a dilution refrigerator. The electronic temperature is calibrated to ~138 mK using the Al(111) surface (The calibration is shown in the Figure S8 of our previous work[52]). All scanning parameters (set point voltage $V_s$ and tunneling current $I_t$) of the STM topographic image are provided in the corresponding figure captions. The $dI/dV$ spectra are acquired using a standard lock-in amplifier with a modulation frequency of 877.1 Hz. Nonmagnetic tungsten tips were fabricated via electrochemical etching and calibrated on a clean Au(111) surface prepared by repeated cycles of sputtering with argon ions and annealing at 500 °C. LEED was employed with a 4-grid detector (*Omicron Spectra LEED*) in the UHV chamber at room temperature.

**Air-stability measurements.** To investigate the air stability, room-temperature STM and Raman spectroscopy were performed on samples exposed to ambient conditions. For STM measurements, after exposure to air, the samples were loaded into a room-temperature UHV-STM system (Omicron, base pressure ~$1\times10^{-10}$ mbar), followed by degassing at about 523 K. Raman spectra were collected under ambient conditions using a customized Raman setup (WITec alpha 300R) with a 532 nm excitation wavelength. High-quality $Nb_2Pd_3Te_5$ bulk crystals used for Raman measurements were obtained through a self-flux method.

**Data availability**

Data measured or analyzed during this study are available from the corresponding authors on request.

**Acknowledgements**

This work is supported by grants from the National Natural Science Foundation of China (62488201, 52572188, 92580202), the National Key Research and Development Projects of China (2022YFA1204100), the CAS Project for Young Scientists in Basic Research (YSBR-053, YSBR-003).

**Author Contributions:** H.-J.G. supervised the project. H.-J.G., H.C., and H.G. designed the experiments. J.W. H.G. S.X., Q.F., H.X. C.M.S. fabricated the samples. H.Z., H.C. and Z.W. performed STM experiments. All the authors participated in analyzing the data, plotting figures, and writing the manuscript.

**Competing Interests:** The authors declare that they have no competing interests.